\documentclass[%
pra,
preprint,
superscriptaddress,
showpacs,
a4paper, 12pt]{revtex4}

\usepackage{amssymb,amsmath}
\usepackage{graphicx}
\usepackage{dcolumn}

\newcommand{\ci}{\mathop\textrm{ci}\nolimits}
\newcommand{\si}{\mathop\textrm{si}\nolimits}

\begin{document}

\title{Effects of interatomic interaction on cooperative relaxation of
two-level atoms}

\author{Sergei~P.~Lukyanets}
\email[Email address: ]{lukyan@iop.kiev.ua}
\affiliation{Institute of Physics of National Academy of Sciences of Ukraine,
 Prospect Nauki 46, 03028 Ky\"{\i}v, Ukraine}
\author{Dmytro~A.~Bevzenko}
\affiliation{Institute of Physics of National Academy of Sciences of Ukraine,
 Prospect Nauki 46, 03028 Ky\"{\i}v, Ukraine}

\date{\today}

\begin{abstract}
   We study effects of direct interatomic interaction on cooperative
  processes in atom-photon dynamics.  Using a model of two-level atoms
  with Ising-type interaction as an example, it is demonstrated that
  interparticle interaction combined with atom-field coupling can
  introduce additional interatomic correlations acting as a phase
  synchronizing factor.  For the case of weakly interacting atoms with
  $J<\hbar\omega_0$, where $J$ is the interparticle coupling constant
  and $\omega_0$ is the atomic frequency, dynamical regimes of
  cooperative relaxation of atoms are analyzed in the Born-Markov
  approximation both numerically and using the mean field
  approximation.  We show that interparticle correlations induced by
  the direct interaction result in inhibition of incoherent
  spontaneous decay leading to the regime of collective pulse
  relaxation which differs from superradiance in nature.  For
  superradiant transition, the synchronizing effect of interatomic
  interaction is found to manifest itself in enhancement of
  superradiance.  When the interaction is strong and
  $J>\hbar\omega_0$, one-particle one-photon transitions are excluded
  and transition to the regime of multiphoton relaxation occurs.
  Using a simple model of two atoms in a high-Q single mode cavity we
  show that such transition is accompanied by Rabi oscillations
  involving many-atom multiphoton states. Dephasing effect of
  dipole-dipole interaction and solitonic mechanism of relaxation are
  discussed.
\end{abstract}



\maketitle

\section{Introduction}
\label{sec:introduction}

The phenomenon of superradiance has a long history dating back
more than 60 years to the seminal paper by Dicke~\cite{Dicke}
where the effect was predicted theoretically. Over the past few
decades the superradiance has been the subject of intense
theoretical and experimental studies in a large variety of
systems. These include molecular
aggregates~\cite{Meinardi,Lim,Lippitz,vanHalst}, cold
atoms~\cite{Kuga, Yelin} and Bose-Einstein
condensates~\cite{Inouye,Schneble1,Schneble2}, atomic
nuclei~\cite{Baldwin,Yukalov,DAN}, magnetic
nanoclusters~\cite{Tejada,Kaganov,Chudnovsky},
heterostructures~\cite{Bardot,Laikhtman,Brandes} and many others.

The key process underlying the mechanism of superradiance is
phase synchronization of initially independent atoms caused by the
coupling with a common environment represented by the
electromagnetic field. In order for such process to occur the
phase decoherence time of atoms should be longer than the photon
travel time in the sample~\cite{Haroche,Menshikov}.

For samples which size is smaller than the wavelength of
radiation, this condition requires  the density of atoms to be
sufficiently high.
 The system-environment (atom-field) coupling also manifests itself as an
additional indirect interaction (a sort of the transverse
dipole-dipole interaction) which may suppress superradiant
transitions depending on the spatial distribution of atoms or the
sample geometry~\cite{Menshikov, Haroche}.

From the other hand, when the density of atoms (or, more
generally, emitters) is high the direct interparticle interaction
starts to play an increasingly important part in determining
cooperative behavior of the particles. In particular, this
interaction strongly affects the properties of  low-dimensional
systems. The Mott-insulator quantum phase transition in optically
trapped atomic systems~\cite{Greiner1, Kasevich, Morsch, Illuminati} and in solid
structures~\cite{Mott, Kittel, Emry, Imada}, generation of many-particle entangled
states or many-particle coherent dynamics, as it is in the case of
effectively interacting atoms inside a high quality dissipative
cavity~\cite{Gabris, Unanyan}, Bose-Einstein
condensate~\cite{Meystre, Kasevich} and in molecular clusters with
strong magnetic~\cite{Loss-nature, Chudnovsky-Tejada} or
Coulomb \cite{Lukyanets, Lukyanets2} correlations are
examples.

The direct interparticle interaction introduces additional
correlations between emitters. These correlations considerably
influence the cooperative optical properties of atoms.

Firstly, the interaction directly affects the superradiance
leading to a number of peculiarities such as changing the order of
superradiant phase transitions~\cite{Johnson,Bogolubov}.
Recently, the possibility of superradiant relaxation in strongly
correlated systems was studied
theoretically~\cite{Chudnovsky,Kaganov}. The experimental results
for magnetic molecules of Mn$_{12}$-ac type were also reported in
Ref.~\cite{Tejada}. Earlier, systems of ferroelectric type with
strong interparticle interaction were regarded as promising
candidates for an active medium of the heat pumping
laser~\cite{Bogolubov}.

Secondly, the direct interparticle interaction can play the role
of a phase synchronizing factor that may lead to the cooperative
behavior which, though shares many common properties with the
superradiance effect, essentially differs  from superradiance in
nature. The classical example furnishes the spectrum of P
 luminescence band in CdS and ZnO where
the emission intensity is proportional to the
second power of the free exciton number (pumping intensity).
In this case the effect is caused by exciton-exciton
scattering~\cite{Salvan,Hvam}. Recently, such effects were
observed in the microcrystalline phase of CsPbCl$_3$ thin
films~\cite{Kondo}.

For interacting atoms, the interaction can drastically change the
regime of atom-photon dynamics by inducing (otherwise, excluded)
multiphoton transitions~\cite{Varada}. It was shown in
Ref.~\cite{Hettich} that interatomic interaction can give rise to
non-zero multiphoton emission observed with single-molecule
spectroscopy technique as a two-photon cooperative effect for
strongly dipole-dipole coupled molecules. Theoretically, this
phenomenon was predicted as a large two-atom two-photon resonant
effect for two atoms inside a high-quality
cavity~\cite{Agarwal04}.

So, different regimes of radiative decay in correlated atomic
systems are mainly governed by the interatomic interaction. By
controlling the interaction radiation properties of such systems
can be widely varied ranging from superradiant transitions to the
generation of the Fock state of light. In particular, such control
is feasible for the atoms in optical lattices (see,
e.g.,~\cite{Greiner1, Greiner2}).

In this work cooperative radiation of interacting atoms coupled to
electromagnetic bath will be of our primary interest. We are aimed
to study different relaxation regimes determined by the intensity
of interatomic coupling.

The paper is organized as follows.

In Sec.~\ref{sec:model} we formulate the model of $N$ two-level
atoms with Ising-type interaction and qualitatively discuss
various regimes of relaxation by considering realignment of the
atomic energy spectrum at different values of the interatomic
coupling constant $J$. There are two limiting cases of weak and
strong interaction with $J<\hbar\omega_0$ and  $J>\hbar\omega_0$,
respectively ($\omega_0$ is the atomic frequency).

We find that, for weakly interacting atoms, Ising interaction
would affect dynamical behavior of the system leading to the
transition to collective pulse relaxation and enhancement of
superradiance. For strong interaction, the regime of multiphoton
relaxation is predicted to occur. It is also shown that dependence
of the radiation intensity peak on the number of particles can be
anomalous at long-range interatomic interaction.

Derivation of the master equation for weakly interacting atoms is
presented in Sec.~\ref{sec:master-equation}. We show that, for
certain atomic configurations, dephasing effects of induced
dipole-dipole interaction can be suppressed and dynamics of the
atomic system can be described by the simplified master equation.

In Sec.~\ref{sec:coop-rad-weakly}, the effects for weakly
interacting atoms briefly discussed in Sec.~\ref{sec:model} are
investigated in detail. By applying the mean field approximation
we obtain the results that agree very well with those calculated
by solving the equations for atomic variables numerically.

The regime of multiphoton relaxation that takes place at strong
interatomic interaction due to inhibition of one-particle one-photon
transitions is described in Sec.~\ref{sec:multiphoton}.  Finally, in
Sec.~\ref{sec:disc-concl}, we draw together the results,
discuss solitonic mechanism of relaxation and make some
concluding remarks.

Details on some technical results are  relegated to
Appendixes~\ref{Ap-Born-Markov}-~\ref{Ap-fast-var}.

\section{Atomic energy spectrum and regimes of relaxation}
\label{sec:model}

In order to illustrate a possibility of different relaxational
regimes, caused by direct interatomic interaction, we consider a
simplest model of a chain of two-level atoms with nearest neighbor
Ising-type interaction. Typically, the models of this type are used to
describe molecular systems with Coulomb and magnetic interactions such
as ferroelectric~\cite{Blinc} and
magnetic~\cite{Loss, Tyablikov, Sachdev} clusters, interacting
electrons in the tightly-binding approximation~\cite{Emry},
interacting atoms in optical lattice~\cite{Illuminati}, where the
extended Hubbard model is usually introduced.

The interaction takes into account repulsion of the neighboring
atoms so that the system may reveal the correlated many-particle
behavior. When the energies of the neighbors are different
tunneling transitions between neighboring sites are inelastic.

In addition to the Ising-type direct interaction, the atoms
interact with a common electromagnetic field (the atom-field
interaction). These two interactions are combined to maintain
correlations in the atomic subsystem thus crucially affecting the
regimes of cooperative optical transitions.

\subsection{The model}
\label{subsec:model-1}

The Hamiltonian of the model atomic chain with the short-range
Ising-type interaction can be written in the following form

\begin{equation}\label{Ham}
H=H_A+H_F+H_I,
\end{equation}
where
\begin{equation}\label{H_atom}
H_A=\hbar\omega_0\sum_{i=1}^N S_i^z - J\sum_{i=1}^N S_i^z
S_{i+1}^z
\end{equation}
is the Hamiltonian of atomic subsystem,
\begin{equation}\label{HField}
H_F=\sum_{\mathbf{k}s}\hbar\omega_k a_{\mathbf{k}s}^+
a_{\mathbf{k}s}
\end{equation}
is the Hamiltonian of electromagnetic field and
\begin{equation}\label{H_int}
H_I=-i\hbar\sum_{i=1}^N\sum_{\mathbf{k}s}\left[g_{\mathbf{k}s,i}\left(S_i^+
+ S_i^-\right)a_{\mathbf{k}s}-H.c.\right]
\end{equation}
is the Hamiltonian of the atom-field interaction and H.c. stands
for Hermitian conjugation; $\omega_0$ is the frequency of atomic
transition, $J>0$ is the coupling constant of the nearest neighbor
interaction, $N$ is the number of atoms, $S_i^z$ is the
$z$-component of the pseudo-spin operator determining the
population of the $i$th atom, $S_i^+$ ($S_i^-$) is the raising
(lowering) operator of the pseudo-spin, $a_{\mathbf{k}s}$
($a_{\mathbf{k}s}^+$) is the photon annihilation (creation)
operator characterized by the wavenumber vector $\mathbf{k}$, the
frequency $\omega_k$ and the polarization vector
$\mathbf{e}_{\mathbf{k}s}$,
\begin{equation}
g_{\mathbf{k}s,i}=\sqrt{\frac{\omega_k}{2\varepsilon_0\hbar{v}}}
\left(\mathbf{d}_i\cdot\mathbf{e}_{\mathbf{k}s}\right){e}^{i\mathbf{kr}_i}
\end{equation}
is the coupling constant of the dipole interaction between the
$i$th atom and transverse electromagnetic field, $\mathbf{d}_i$ is
the effective dipole moment of $i$th two-level atom with the
vector of spatial coordinates $\mathbf{r}_i$, $\varepsilon_0$ is
the vacuum dielectric constant and $v$ is the volume of field
quantization.

The Hamiltonian~\eqref{Ham} is distinguished from standard models
of two-level atom interacting with the electromagnetic field by
the presence of the interatomic interaction in $H_A$ [see
Eq.~\eqref{H_atom}].

In what follows the dynamics of the collective relaxation of
initially inverted atomic subsystem will be of our primary
concern. More specifically, we shall study various dynamical
scenarios that occur at different values of the coupling constant
$J$. In the remaining part of this section we first discuss energy
spectrum of the atomic subsystem and qualitatively describe these
dynamical regimes.

\subsection{Weak interatomic interaction, $J<\hbar\omega_0$}
\label{sec:weak-interaction}

The transition energy of an atom depends on the states of its
neighbors as a result of interatomic interaction.
This effect can be described as renormalization of the
transition frequency of the $i$th atom
$\omega_0\rightarrow\tilde\omega_i$ which can also be readily seen
from the mean field expression for the Hamiltonian $H_A$:
\begin{equation*}
H_A\sim\sum_i\hbar\left[\omega_0-\frac{J}{2\hbar}\left(\langle
S^z_{i-1}\rangle + \langle S^z_{i+1}\rangle\right)\right]
S^z_i=\sum_i\hbar\tilde\omega_i S^z_i.
\end{equation*}

\begin{figure*}[!tbh]
 \centering
 \resizebox{100mm}{!}{\includegraphics{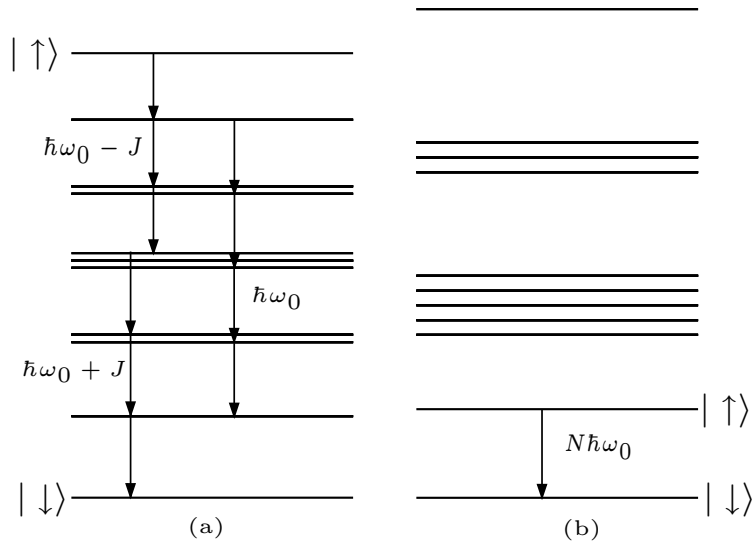}}
  \caption{Energy spectrum of $H_A$ for the chain of six atoms.
Two cases are shown: (a) weak interatomic interaction with
   $J/\hbar\omega_0=0.1$ and (b) strong interatomic interaction with
   $J/\hbar\omega_0=10$.
 The state
$\vert\uparrow\rangle=\vert\uparrow,\uparrow,...,
\uparrow\rangle$ ($\vert\downarrow\rangle=\vert\downarrow,\downarrow,...,\downarrow\rangle$)
indicates that the atoms are all in the excited (ground) state.}
  \label{spectra}
\end{figure*}

Sufficiently weak interaction, $J\ll\hbar\omega_0$, results in inhomogeneous
broadening of the radiation spectrum.
When the coupling constant of interatomic interaction $J$
increases and $J<\hbar\omega_0$, the spectrum of the atomic
subsystem undergoes more pronounced rearrangement. As is shown in
Fig.~\ref{spectra}, the spectrum is no longer equidistant, but its
structure is identical to that for non-interacting atoms with
$J=0$. In particular, the latter implies that the state with all
the atoms excited $|\uparrow\rangle=|\uparrow, \uparrow, ...,
\uparrow\rangle$ (spin up represents the excited state of the
atom) gives the level of highest energy and the level of lowest
energy corresponds to the case where atoms are all in the ground
states $|\downarrow\rangle=|\downarrow, \downarrow, ...,
\downarrow\rangle$.

Similar to the case of non-interacting atoms, the dynamics of
radiative decay is dominated by one-particle one-photon
transitions and can be described using the Born-Markov
approximation.

A cooperative relaxation of the system from an excited state can
be realized by different schemes of transitions. It is a possibility to
have a generation in the two mode regime with the frequencies
$\omega=\omega_0\pm J/\hbar$ (see Fig.~\ref{spectra}a) arising due
to renormalization of the atomic frequency $\tilde\omega_i$ that
governs the intensity of radiation.

It is well known~\cite{Menshikov} that the intensity of atom
radiation $I$ is proportional to the fourth power of the transition
frequency $I\propto\omega_0^4|\mathbf{d}|^2$, where $\mathbf{d}$ is
the transverse dipole moment of the atomic transition. For $N$
weakly interacting atoms, the intensity $I$ is defined by the
renormalized frequency and can be represented by a sum of the
coherent and the incoherent parts, $I=I_\textrm{coh}+I_\textrm{incoh}$, where
$I_\textrm{coh}\propto\sum_{i\neq
  j}\tilde{\omega}_i^4|\mathbf{d}_i\mathbf{d}_j^*|\propto\tilde\omega^4N^2$
and
$I_\textrm{incoh}\propto\sum_i\tilde{\omega}_i^4\mathbf{d}_i^2\propto\tilde\omega^4N$.

Dependence of the renormalized frequency on the states of
neighboring atoms introduces additional correlations into the
incoherent part of radiation. Such correlations may result in a
cooperative radiation of fundamentally different origin than
superradiance described by the coherent part $I_\textrm{coh}$.

The effect of superradiance, for its part, can be enhanced by the
Ising-type interaction which may act as a phase synchronizing
factor. It is also worth noting that, for a long-range
interparticle interaction, the renormalized frequency $\tilde\omega_i$ depends on
the number of atoms, $\tilde\omega_i\sim\omega_0\left(1-\frac{1}{2}\sum_{j\neq
    i}\frac{J_{ij}}{\hbar\omega_0}\langle S^z_j\rangle\right)$, leading to anomaly in the $N$-dependence of the intensity peak of
collective radiation.

Figure~\ref{spectra}a shows that, when the atomic subsystem is not
completely inverted, an excited state may relax through a series
of transitions which frequencies are identically equal to the
resonant frequency $\omega_0$. This regime of relaxation
resembles the solitonic mechanism where the process of emission is
connected with inelastic motion of defects (Bloch walls) induced
by a short-range interaction. We shall discuss this point at
greater length in Sec.~\ref{sec:disc-concl}.

\subsection{Strong interatomic interaction, $J>\hbar\omega_0$}
\label{sec:strong-interaction}

In the case of strong interaction with $J>\hbar\omega_0$,
arrangement of the energy level significantly differs from that of
non-interacting atoms. Now the state $|\uparrow\rangle$ does not
give the level of highest energy and $E_N<E_{N-1}$, where $E_j$ is
the energy of the state with $j$ excited atoms. Such realignment
indicates that many-atom multiphoton transitions will dominate the
process of relaxation.

Figure~\ref{spectra}b illustrates that, in the limiting case with
$J\gg\hbar\omega_0$, the states $|\uparrow\rangle$ and
$|\downarrow\rangle$ give two lowest lying energy levels. They are
separated from the other excited states by the gap of the width
$\sim J$. Relaxation of fully inverted state (transition from
$|\uparrow\rangle$ to $|\downarrow\rangle$) can be achieved
through a $N$-photon process.

At this stage it should be stressed that an accurate description
of the interaction between the system and the field in the
presence of strong interatomic correlations requires careful
consideration of interplay between many-atom multiphoton dipole
and multipole transitions. In a sense, our system can be regarded
as an intermediate case between a collection of two-level atoms
independently interacting with the field and a single many-level
system.

A good example provides the spin-phonon interaction in magnetic
molecules Mn$_{12}$O$_{12}$ of the multipole form
$H_{I}\propto(S^+)^4a_k+(S^-)^4a^+_k$ which symmetry is consistent
with the anisotropy energy~\cite{Politi}. Similarly, the spin
superradiance in atomic nuclei can be dominated by multipole
transitions~\cite{DAN}.

For the multipole superradiance~\cite{Kopvillem}, the exponent of
the $N$-dependence of the radiation intensity peak ($N$ is the
number of emitters) can be greater than two, $I\propto N^{\alpha}$
with $\alpha>2$. This is the case for the cooperative spin-phonon
relaxation in Mn$_{12}$O$_{12}$ where a rough estimate gives
$I\propto\langle(S^+)^4(S^-)^4\rangle\propto N^8$.

The standard approach to the dissipative dynamics of atoms relies
on the Lindblad equation for the reduced density matrix of the
atomic subsystem derived by eliminating boson variables from the
master equation for the density operator of atom-field system. This
master equation is obtained using the Born-Markov
approximation~\cite{Carmichael} that neglects multiphoton
transitions and thus is applicable only for weakly interacting
atoms at $J<\hbar\omega_0$. In the subsequent section
we discuss the derivation procedure for the Lindblad equation for the
density matrix of weakly interacting atoms.

\section{Master equation for weakly interacting atoms}
\label{sec:master-equation}

In this section we derive and discuss the master equation for the
reduced atomic density operator $\rho$ for the case of weak
interaction at $J<\hbar\omega_0$. We show that the direct
interaction renormalizes both the induced dipole-dipole
interaction and the damping operator in the Lindblad equation. In
addition, we find that, under certain conditions, destructive
effect of the dipole-dipole interaction on supperradiance is
suppressed.

\subsection{Derivation of master equation}
\label{subsec:deriv-mast-equat}

We begin by applying the standard procedure of  elimination of
field variables in the Born-Markov approximation~\cite{Carmichael}
to the equation for the density matrix of the combined field-atom
system, $\rho_{AF}$, written in the representation of interaction
\begin{equation}
\label{ME_gen_form}
i\hbar\frac{d}{dt}\tilde{\rho}_{AF}(t)=\left[V(t),\tilde{\rho}_{AF}(t)\right],
\end{equation}
where
\begin{eqnarray}
\tilde{\rho}_{AF}(t)&=&{e}^{i(H_A+H_F)t/\hbar}\rho_{AF}{e}^{-i(H_A+H_F)t/\hbar},\nonumber\\
V(t)&\equiv&{e}^{i(H_A+H_F)t/\hbar}H_I{e}^{-i(H_A+H_F)t/\hbar}=\nonumber\\
&=&-i\hbar
\sum_{i=1}^N\sum_{\mathbf{k}s}\left[g_{\mathbf{k}s,i}\left(S_i^+e^{i\left(\omega_0\hat{\Gamma}_i-\omega_k\right)t}
+
S_i^-e^{-i\left(\omega_0\hat{\Gamma}_i+\omega_k\right)t}\right)a_{\mathbf{k}s}-H.c.\right],
\label{V_int}\\
\hat{\Gamma}_i&=&1-\beta\left(S_{i+1}^z+S_{i-1}^z\right),
~~~\beta=J/(\hbar\omega_0) \label{Gamma}
\end{eqnarray}
and the expression for the operator~\eqref{V_int} is derived using
the cyclic boundary conditions.

As it was pointed out in Sec.~\ref{sec:model}, for the $i$th atom,
the energy of transition depends on the state of the neighboring
atoms. Alignment of the energy spectrum of the atomic subsystem
and, as a result, the character of atom-field interaction are
determined by the value of the ratio  $\beta=J/\hbar\omega_0$.
When $J>\hbar\omega_0$ and $\beta>1$, ordering of the energy
levels changes and certain transitions appear to be excluded.

In Eq.~\eqref{V_int} the frequency of the dipole transition is
effectively described by the operator~\eqref{Gamma}
$\omega_0\hat{\Gamma}_i$ entering the Hamiltonian of interaction
$V$. As opposed to the case with $\beta<1$, the eigenvalues of the
operator $\hat\Gamma_i$ can be negative at $\beta>1$ depending on
$\beta$ and $i$.

In this case, for example, the energy $E_N$ of the ``monodomain''
state $|\uparrow, \uparrow, ..., \uparrow\rangle$ is below the
energy $E_{N-1}$ of the state where only one atom is in the ground
state $|\downarrow, \uparrow,\uparrow, .. ,\uparrow\rangle$. The
result is that relaxation of the completely inverted state
 $|\uparrow, \uparrow, ..., \uparrow\rangle$
cannot occur as a one-photon process.

It should be noted that, for $\beta>1$, the change of sign in the
eigenvalues of the operator $\hat\Gamma_i$ introduces new resonant
terms of the form $S^+_ia^+_{\mathbf{k}s}$. However, as it was
discussed in Sec.~\ref{sec:strong-interaction}, this case is
additionally complicated by multipole transitions that need to be
taken into consideration.

When $J<\hbar\omega_0$ and $\beta<1$, the eigenvalues of the
operator $\hat\Gamma_i$ are all positive. So the Ising-type
interaction just renormalizes the transition frequencies inducing
additional correlations in the atomic subsystem.

Dynamics of the atomic relaxation is now governed by one-photon
processes and, in the remaining part of this section, we restrict
ourselves to this case of weakly interacting atoms with $\beta<1$.

We can now use the Born approximation to derive the equation for
the reduced density matrix of the atomic
subsystem~\cite{Carmichael}, $\rho=Tr_F\{\rho_{AF}\}$. For
simplicity, we consider the zero-temperature case where only
vacuum fluctuations of the electromagnetic field are taken into
account. So, the density operator $\rho_{AF}$ can be written in
the following form
\begin{equation}
\label{DM_B} \rho_{AF}(t)=\vert 0\rangle\langle
0\vert\otimes\rho(t).
\end{equation}

We additionally assume that the wavelength of radiation is much
longer than the size of the system and use the Markov
approximation~\cite{Haroche}. Omitting the technical details that
can be found in Appendix~\ref{Ap-Born-Markov}, the resulting
equation is given by
\begin{eqnarray}
\label{ME_B-M_1}
\frac{d}{dt}\tilde{\rho}(t)&=&\eta\sum_{i,j}\int_0^\infty{d\omega}\omega^3
{F_{ij}(\omega)}
\Biggl\{[R_j(t),\tilde{\rho}(t)S_i^+(t)(\pi\delta(\omega-\omega_0\hat{\Gamma}_i)
+\frac{i\mathcal{P}}{\omega-{\omega_0}
\hat{\Gamma_i}})]+
\nonumber\\
&+&
[R_j(t),\tilde{\rho}(t)S_i^-(t)\frac{i\mathcal{P}}{\omega+{\omega_0}
\hat{\Gamma_i}}]+H.c.\Biggr\},
\end{eqnarray}
where
\begin{eqnarray}
S_i^\pm(t)&\equiv&{e}^{i(H_A+H_F)t/\hbar}S_i^\pm{e}^{-i(H_A+H_F)t/\hbar}={S_i^\pm}e^{\pm{i\omega_0}\hat{\Gamma}_it},
\nonumber\\
\tilde{\rho}(t)&=&{e}^{i(H_A+H_F)t/\hbar}\rho{e}^{-i(H_A+H_F)t/\hbar},
\label{rho-int}\\
R_i(t)&=&S_i^+(t)+S_i^-(t),
\nonumber
\end{eqnarray}
and
\begin{equation}
\label{F}
F_{ij}(\omega)=\frac{3}{2}\left\{[1-(\bar{\mathbf{d}}\cdot\bar{\mathbf{r}}_{ij})^2]\frac{\sin{kr_{ij}}}{kr_{ij}}+
[1-3(\bar{\mathbf{d}}\cdot\bar{\mathbf{r}}_{ij})^2]\left(\frac{\cos{kr_{ij}}}{(kr_{ij})^2}-\frac{\sin{kr_{ij}}}{(kr_{ij})^3}\right)\right\},
\end{equation}
$k=\omega/c$, $r_{ij}\equiv|\mathbf{r}_{ij}|=|\mathbf{r}_{i}-\mathbf{r}_{j}|$ is the separation
distance between $i$th and $j$the atoms, we assume $\mathbf{d}_i=\mathbf{d}_j\equiv \mathbf{d}$,
$\mathbf{\bar{r}}_{ij}\equiv\mathbf{r}_{ij}/|\mathbf{r}_{ij}|$,  $\mathbf{\bar{d}}\equiv\mathbf{d}/|\mathbf{d}|$ and
\begin{equation}
\label{eta} \eta=\frac{d^2}{6\varepsilon_0\hbar\pi^2c^3}.
\end{equation}

For brevity, in Eq.~\eqref{ME_B-M_1} we used operator functions
defined on eigenvalues of $\hat{\Gamma}_i$, so that if
$\hat{\Gamma}_i|\psi\rangle=\Gamma_i|\psi\rangle$ then, e.g.,
$\delta(\omega-\omega_0\hat{\Gamma}_i)|\psi\rangle=|\psi\rangle\delta(\omega-\omega_0\Gamma_i)$.
It should be stressed once again that Eq.~\eqref{ME_B-M_1} is
valid only for the case of weakly interacting atoms, when
$J<\hbar\omega_0$ and the Born-Markov approximation is applicable.

Principal values of the integrals that enter the right hand side
of the master equation~\eqref{ME_B-M_1} define the dipole-dipole
interatomic interaction (see Appendix~\ref{Ap-Born-Markov}) and the Bethe-part of the Lamb shift, which
depend on the direct interparticle interaction.
Equation~\eqref{ME_B-M_1} can be further simplified under the condition
that the characteristic wavelength of radiation $\lambda$ is much
longer than the size of the sample.

It is well
known~\cite{Haroche} that in this case the dipole-dipole
interaction is of quasi-static character and is independent of the
frequency of atomic transition.
Since the direct interaction in Eq.~\eqref{V_int} just
renormalizes the atomic frequencies
$\omega_0\rightarrow\tilde{\omega}_i=\omega_0\hat{\Gamma}_i$, the
asymptotic form of the dipole-dipole interaction in the limit
$r_{ij}/\lambda\ll1$ must be identical to that for the system of
non-interacting atoms with $J=0$. At the same time the relaxation
terms proportional to $\tilde{\omega}_i^3$ essentially depend on
the interatomic interaction.

Mathematically, Eq.~\eqref{d-d-accur} in the limit
$r_{ij}/\lambda\ll1$ gives the relation
\begin{eqnarray}
\label{d-d-as}
\mathcal{P}\int_0^\infty{d\omega}\omega^3\frac{F_{ij}(\omega)}{\omega\pm\omega_0\hat{\Gamma}_i}\sim-
\frac{3}{4}\frac{\pi}{(r_{ij}/c)^3}\left[1-3(\mathbf{d}\cdot\hat{\mathbf{r}}_{ij})^2\right],~~~i\neq j.
\end{eqnarray}
With $F_{ij}(\omega)\approx1$ and the Lamb shift
disregarded Eq.~\eqref{ME_B-M_1} gives the master equation in the
simplified form
\begin{equation}
\label{ME_B-M_2} \frac{d}{dt}\tilde{\rho}(t)=\gamma_0\sum_{i,j}
[R_j(t),\tilde{\rho}(t)\hat{\Gamma}_i^3S_i^+(t)]
+i\sum_{i\neq{j}}
\frac{\Omega_{ij}}{2}[R_j(t),\tilde{\rho}(t)R_i(t)]+H.c.,
\end{equation}
where $\gamma_0=\pi\eta\omega_0^3$ is one-half the spontaneous decay rate
for an isolated atom and
\begin{equation}
\label{omega}
\Omega_{ij}=-\pi\eta\frac{3}{2}\cdot\frac{1-3(\mathbf{d}\cdot\hat{\mathbf{r}}_{ij})^2}{(r_{ij}/c)^3}
\end{equation}
is the constant of the quasi-static dipole-dipole interaction~\cite{Haroche}.

The first term on the right hand side of Eq.~\eqref{ME_B-M_2}
describes renormalization of the damping term by the interatomic
interaction. After neglecting the rapidly oscillating terms (the rotating wave approximation) in
Eq.~\eqref{ME_B-M_2} and going back to the Schr\"odinger
representation, we derive the master equation for the case of
weakly interacting atoms, $\beta<1$, in the final form:
\begin{eqnarray}
\frac{d}{dt}\rho(t)&=&-\frac{i}{\hbar}[H_A,\rho(t)]+i\sum_{i\neq{j}}\Omega_{ij}[S_i^+
S_j^-,\rho(t)]+
\nonumber\\
&+&\gamma_0\sum_{i,j}\left([S_j^-,\rho(t)\hat{\Gamma}_i^3S_i^+]+
[S_i^-\hat{\Gamma}_i^3\rho(t),S_j^+]\right).
\label{ME1}
\end{eqnarray}

Eq.~\eqref{ME1} describes dynamics of the Ising-like interacting
atoms coupled to electromagnetic bath. The key feature of this
equation is that only the damping terms are renormalized by the
direct interatomic interaction at $r_{ij}\ll\lambda$.

\subsection{Effects of dipole-dipole interaction}
\label{subsec:dipole-dipole}

It is known (see, e.g., reviews~\cite{Haroche, Menshikov}) that the
dipole-dipole interaction may destroy the cooperative radiation
due to the frequency chirping effect. This effect is caused by the
dynamical shift of energy levels that results in a rapid loss of
phase synchronization with the characteristic time of phase
decoherence much shorter than the rate of relaxation~\cite{Haroche}.

Influence of the dipole-dipole interaction on the dynamics of the
atomic subsystem is strongly affected by the spatial configuration
of atoms and the angular distribution of dipole moments. It turns
out that, for certain configurations, dephasing induced by the
dipole-dipole interaction is inhibited and the cooperative
radiation takes place~\cite{Haroche, Menshikov}.

In particular, the latter is the case when atoms are all in
identical environment and the system is  invariant under
permutations of atoms, $\Omega_{ij}=\Omega_{i'j'}=\Omega$. These
conditions are met for a two atom system or an atomic
ring~\cite{Haroche}.

For disordered atomic configuration (atomic cloud), the
superradiance effect depends on the sample shape. It can occur for pencil-shape patterns as a result of
suppression of the destructive effect of the dipole-dipole
interaction~\cite{Menshikov}.

In our case, when the interatomic spacings is much smaller than
the wavelength ($r_{ij}\ll\lambda$), the form of the dipole-dipole
interaction~\eqref{omega} is the same for both interacting and
non-interacting atoms. Symmetry of the renormalized damping terms
is also identical to the symmetry of the system of non-interacting
atoms with $J=0$. So, atomic configurations suppressing the
dephasing effect of the dipole-dipole interaction are the same for
both cases: $J=0$ and $J\ne 0$.

For the simplest case of two atomic configuration, this point can
be illustrated explicitly. Similar to the case of two
non-interacting atoms considered in Ref.~\cite{Coffey}, the master
equation~\eqref{ME1} for two atoms with $J\ne 0$ is exactly
solvable. The exact solution described in Appendix~\ref{Ap-2atBM}
shows that relaxation is independent of the dipole-dipole
interaction.

The analytical results presented in Appendix~\ref{Ap-2atBM} can
also be used to demonstrate the effect of additional correlations
induced by the interaction. For this purpose, we consider the
expression for the relaxation rate of initially excited atoms
\begin{equation}
\label{2atdec_rate}
\gamma(t)\equiv-\frac{d}{dt}{Tr}\lbrace\rho(S_1^z+S_2^z)\rbrace=4me^{-4mt}\left[1+\frac{n}{n-m}\left(1-e^{-4(n-m)t}\right)\right],
\end{equation}
where $n=\gamma_0(1+\beta/2)^3$ and $m=\gamma_0(1-\beta/2)^3$.

\begin{figure*}[!tbh]
\centering
 \resizebox{100mm}{!}{\includegraphics{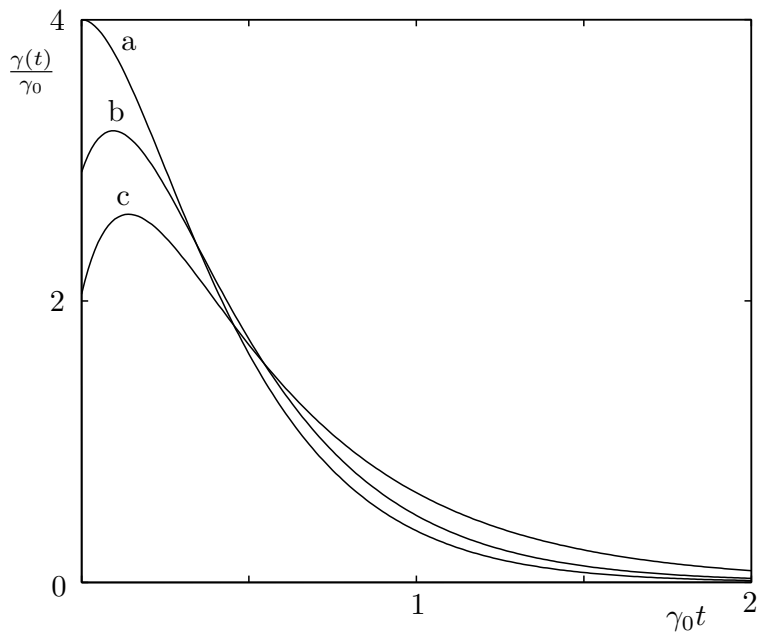}}
 \caption{Relaxation rate $\gamma(t)$
   \eqref{2atdec_rate} for two interacting two-level atoms
at various values of the ratio $\beta\equiv J/\hbar\omega_0$:
(a)~$\beta=0$, (b)~$\beta=0.1$, and (c)~$\beta=0.2$.}
  \label{2atBM}
\end{figure*}

In Fig.~\ref{2atBM} the relaxation rate~\eqref{2atdec_rate} is
plotted as a function of time at various values of $\beta$. It is
seen that the interaction noticeably influences the character of
relaxation even at small values of $\beta$. By contrast to the
case of non-interacting atoms with $\beta=0$, the system features
collective decay with a pronounced emitted pulse arising as a
result of additional correlations induced by Ising-type
interaction.

Now it is our primary task to examine how the cooperative
relaxation is influenced by the interatomic interaction. So, in
what follows we shall restrict our study to the atomic
configurations where the frequency chirping effect can be
disregarded. For such configurations, the master
equation~\eqref{ME1} can be considerably simplified by neglecting
the terms describing the dipole-dipole interaction.

\section{Cooperative radiation of weakly interacting atoms}
\label{sec:coop-rad-weakly}

From the master equation~\eqref{ME1} without the dipole-dipole
interaction, $\Omega_{ij}=0$, it is straightforward to deduce a
system of equations for averages of the atomic variables $\langle
S^z_n\rangle=Tr_A\{\rho S^z_n\}$ and $\langle
S^\pm_n\rangle=Tr_A\{\rho S^\pm_n\}$. The result is given by

\begin{subequations}
\label{Bloch}
\begin{eqnarray}
\frac{d}{dt}\langle
S^z_n\rangle&=&-\gamma_0\sum_i\langle\hat\Gamma_i^3S_i^+S^-_n+
S^+_nS^-_i\hat\Gamma^3_i\rangle, \label{Bloch_a}
\\
\frac{d}{dt}\langle S^+_n\rangle&=&i\omega_0\langle
S^+_n\hat\Gamma_n\rangle+2\gamma_0\sum_i\langle\hat\Gamma_i^3S_i^+
S^z_n\rangle, \label{Bloch_b}
\\
\frac{d}{dt}\langle S^-_n\rangle&=&-i\omega_0\langle
S^-_n\hat\Gamma_n\rangle+2\gamma_0\sum_i\langle S^z_nS_i^-
\hat\Gamma_i^3\rangle, \label{Bloch_c}
\end{eqnarray}
\end{subequations}
where $\gamma_0=\pi\eta\omega_0$ is one-half the spontaneous
decay rate for an isolated atom.

The relaxation rate of the atomic subsystem
\begin{equation}
\label{decay-rate} \gamma(t)=-\sum_n\frac{d\langle
S^z_n\rangle}{dt}=\gamma_0\sum_{n,i}\langle\hat\Gamma_i^3S_i^+S^-_n+S^+_nS^-_i\hat\Gamma^3_i\rangle
\end{equation}
is of primary importance in our subsequent analysis. The parameter
$\gamma(t)$ is defined in the right hand side of
Eq.~\eqref{Bloch_a}.

By contrast to the case of non-interacting atoms, there is no an
intimate connection between the relaxation rate $\gamma(t)$ and
the total intensity of radiation $I(t)$ which is proportional to
$\sum_{i,n}\langle
S^+_nS^-_i\hat\Gamma^4_i+S^+_i\hat\Gamma^4_iS^-_n\rangle$. The
reason is that the interatomic interaction renormalizes the
frequency of the dipole transition
$\tilde\omega\propto\omega_0\hat\Gamma$, whereas
$\dot{S}^z\propto\tilde\omega^3$ and $I\propto\tilde\omega^4$.

The right hand side of Eq.~\eqref{decay-rate} can be conveniently
rewritten as a sum of the coherent ($i\neq n$) and incoherent
($i=n$) parts
\begin{equation}
\label{occup} \frac{d}{dt}\langle
S^z_n\rangle=-\gamma_0\langle\left(1+2S^z_n\right)\hat\Gamma_n^3\rangle-\gamma_0\sum_{i\neq
n}\langle\hat\Gamma_i^3S_i^+S^-_n+S^+_nS^-_i\hat\Gamma^3_i\rangle,
\end{equation}
where we used the identities $S^+_iS^-_i=\frac{1}{2}+S^z_i$ and $[S^z_n,\hat\Gamma_n]=0$.

In order to decouple correlations in Eq.~\eqref{Bloch}, we use the
Bloch representation for the wave functions of two-level atoms
$|\Phi\rangle$~\cite{Allen}
\begin{eqnarray}
\label{B}
&&|\Phi\rangle=\prod_{j=1}^N|\theta_j,\varphi_j\rangle,\nonumber\\
&&|\theta_j,\varphi_j\rangle=\sin\frac{\theta_j}{2}e^{-i\varphi_j/2}|\uparrow\rangle_j+
\cos\frac{\theta_j}{2}e^{i\varphi_j/2}|\downarrow\rangle_j
\end{eqnarray}
and obtain the closed system of equations for the averages of
atomic variables $\langle
S^z_j\rangle=-\frac{1}{2}\cos\theta_j,~~\langle
S^\pm_j\rangle=\frac{1}{2}\sin\theta_je^{\pm i\varphi_j}$ of the
following form:
\begin{eqnarray}
\label{Bloch2} \frac{d}{d\tau}\langle
S^z_n\rangle&=&-\left(1+2\langle
  S^z_n\rangle\right)\langle\hat\Gamma_n^3\rangle-
\left(\langle S^+_{n+1}\rangle\langle S^-_{n}\rangle+\langle
  S^+_{n}\rangle\langle S^-_{n+1}\rangle\right)
\langle\hat K_{n+1}\rangle -\nonumber
\\
&-&\left(\langle S^+_{n-1}\rangle\langle S^-_{n}\rangle+\langle
  S^+_{n}\rangle\langle
S^-_{n-1}\rangle\right)\langle\hat K_{n-1}\rangle -\nonumber
\\
&-&\sum_{i\neq n,n\pm1}\langle\hat\Gamma_i^3\rangle \left(\langle
S^+_{n}\rangle\langle S^-_{i}\rangle+\langle
  S^+_{i}\rangle\langle S^-_{n}\rangle\right),
\nonumber\\
\frac{d}{d\tau}\langle S^\pm_n\rangle&=&\left(\pm
  i\alpha\langle\hat\Gamma_n\rangle-
\langle\hat\Gamma_n^3\rangle\right)\langle
S^\pm_n\rangle+2\langle\hat E_{n+1}\rangle \langle
S_{n+1}^\pm\rangle+ 2\langle\hat E_{n-1}\rangle \langle
S_{n-1}^\pm\rangle+
\nonumber\\
&+&2\sum_{i\neq n,n\pm1}\langle S^z_n\rangle\langle
S^\pm_i\rangle\langle\hat\Gamma_i^3\rangle,
\end{eqnarray}
where $\tau=\gamma_0t$, $\alpha=\omega_0/\gamma_0\gg1$ and
\begin{eqnarray}
\langle\hat\Gamma_i^3\rangle&=& 1-\beta\left(3+\beta^2\right)
\left(\langle S^z_{n-1}\rangle + \langle
S^z_{n+1}\rangle\right)+\frac{3\beta^2}{2} \left(1+4\langle
S^z_{n-1}\rangle\langle S^z_{n+1}\rangle\right),
\\
\langle\hat
K_{n\pm1}\rangle&=&1+\beta\left(3+3\beta+\beta^2\right)
\left(\frac{1}{2}-\langle S^z_{n\pm2}\rangle\right),
\\
\langle\hat
E_{n\pm1}\rangle&=&\langle\hat\Gamma^3_{n\pm1}S^z_n\rangle=\langle
S^z_n\rangle- \beta(3+\beta^2)\left(\frac{1}{4}+ \langle
S^z_{n\pm2}\rangle\langle S^z_n\rangle\right)+
\frac{3\beta^2}{2}(1+\langle S^z_{n\pm2}\rangle).
\end{eqnarray}

There is a rapidly oscillating term
$i\alpha\langle\hat\Gamma_n\rangle\langle S^\pm_n\rangle$ in the
system of equations~\eqref{Bloch2}. In contrast to the case of
non-interacting atoms with $\hat\Gamma_i\equiv1$, this term cannot
be removed by the phase shift $\langle S^\pm_z\rangle\rightarrow
e^{\pm i\alpha \tau}\langle S^\pm_z\rangle$. In
Appendix~\ref{Ap-fast-var} we apply the method of multitime
scales~\cite{Pobedrya,Nayfeh} to eliminate the rapidly oscillating terms
and deduce the following equations for slowly
varying atomic amplitudes
\begin{equation}
\sigma^z_n(\tau)=\overline{(\langle S^z_n\rangle(\tau,\tau'))}_{\tau'}, ~~~
\sigma_n^\pm(\tau)=\overline{\left(\langle S^\pm_n\rangle(\tau,\tau')
\exp\left(\mp
i\int_0^{\tau'}dt''\langle\hat\Gamma_n\rangle(\tau,t'')\right)\right)}_{\tau'},
\end{equation}
where $\overline{(...)}_{\tau'}$ denotes averaging over the time $\tau'=\alpha\tau$ of fast motion,
\begin{eqnarray}
\label{MF_F} \frac{d}{d\tau}
\sigma^z_n&=&-(1+2\sigma^z_n)\Gamma^3_n-\left(\sigma^+_{n+1}\sigma^-_nw_{n+1,n}+
\sigma^+_n\sigma^-_{n+1}w_{n,n+1}\right)K_{n+1}-
\nonumber\\
&-&
\left(\sigma^+_{n-1}\sigma^-_nw_{n-1,n}+\sigma^+_n\sigma^-_{n-1}w_{n,n-1}\right)
K_{n-1}-\sum_{i\neq n,n\pm1}
\Gamma_i^3\left(\sigma^+_n\sigma^-_iw_{n,i}+\sigma^+_i\sigma^-_nw_{i,n}\right),
\nonumber\\
\frac{d}{d\tau} \sigma^+_n&=&-\sigma^+_n\Gamma^3_n+2
E_{n+1}\sigma^+_{n+1}w_{n+1,n}+ 2
E_{n-1}\sigma^+_{n-1}w_{n-1,n}+2\sum_{i\neq
  n,n\pm1}\Gamma_i^3\sigma_i^+ \sigma^z_nw_{i,n},
\nonumber\\
\sigma^-_i&=&(\sigma_i^+)^*,
\end{eqnarray}
where
\[w_{i,j}=-i~\frac{e^{i2\pi(\Gamma_i-\Gamma_j)}-1}{2\pi(\Gamma_i-\Gamma_j)}\]
and functions
\begin{eqnarray}
&&\Gamma_n=1-\beta\left(\sigma^z_{n+1}+\sigma^z_{n-1}\right),
\label{Gamma_av}
\\
&&\Gamma^3_n=1-\beta\left(3+\beta^2\right)\left(\sigma^z_{n+1}+\sigma^z_{n-1}\right)+
\frac{3}{2}\beta^2\left(1+4\sigma^z_{n+1}\sigma^z_{n-1}\right),
\label{Gamma3_av}
\\
&&K_{n\pm1}=1+\beta\left(3+3\beta+\beta^2\right)\left(\frac{1}{2}-\sigma^z_{n\pm2}\right),
\\
&&E_{n\pm1}=\sigma^z_n-\frac{1}{4}\beta\left(3+\beta^2\right)
\left(1+4\sigma^z_{n\pm2}\sigma^z_n\right)+\frac{3}{2}\beta^2\left(1+\sigma^z_{n\pm2}\right)
\end{eqnarray}
describe correlations due to the direct interatomic interaction.

\subsection{Order parameter and superradiance transition}
\label{subsec:order-parameter}

It is known~\cite{Menshikov} that relaxation of the atomic
subsystem essentially depends on the number of atoms $N$. For
non-interacting atoms ($J=0$ and $\hat\Gamma_i\equiv1$), the first
term on the right hand side of Eq.~\eqref{occup} describes
processes of spontaneous incoherent decay.

Transition to the regime of superradiant relaxation, where
correlations are caused by interaction with the electromagnetic
field in the vacuum state, occurs when the term describing the
coherent part in Eq.~\eqref{occup} (a sum with $i\ne n$) start to
play the dominating role. Such transition can be conveniently
described using the order parameter of the form (see,
e.g.~\cite{Temnov}):
\begin{equation}
\label{C} C=\lim_{T\rightarrow\infty} \frac{1}{T} \int_0^Tdt
\frac{\gamma_\textrm{coh}(t)}{\gamma_\textrm{incoh}(t)}=\lim_{T\rightarrow\infty}
\frac{1}{T} \int_0^Tdt \frac{\sum_j^N\sum_{i\neq j}^N
\langle\hat\Gamma_i^3S_i^+S^-_j+S^+_jS^-_i\hat\Gamma^3_i\rangle(t)}
{\sum_{i}^N\langle{S}^+_iS^-_i\hat\Gamma_i^3+\hat\Gamma_i^3{S}^+_iS^-_i\rangle
(t)}
\end{equation}
and takes place if the number of atoms $N$ exceeds its critical
value $N_c$~\cite{Menshikov}.

The presence of the factor
$\hat{\Gamma}_i=1-\beta\left(S_{i+1}^z+S_{i-1}^z\right)$ in
Eq.~\eqref{C} indicates that the direct interaction introduces
additional correlations into both the coherent and the incoherent
parts of radiation. It is the effect of such correlations on the
incoherent part is responsible for the regime of pulse
radiation~\cite{Salvan, Hvam} discussed in
Sec.~\ref{sec:introduction}.

\begin{figure*}[!tbh]
\centering
 \resizebox{100mm}{!}{\includegraphics{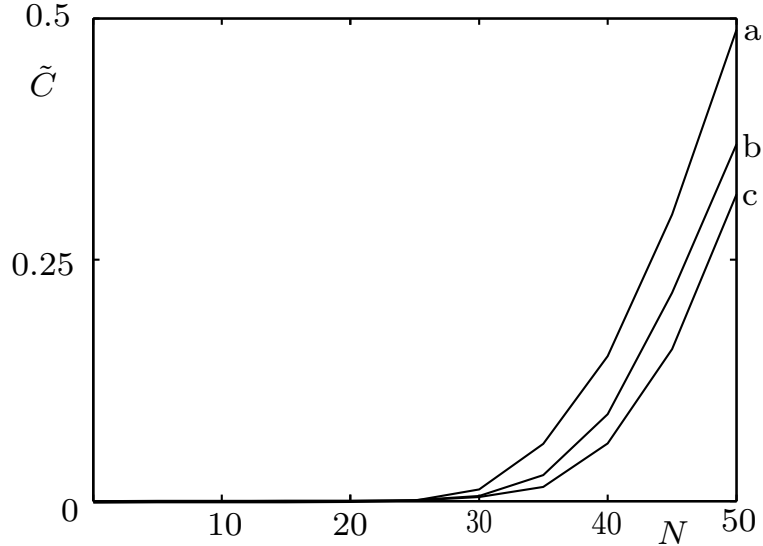}}
 \caption{ Order parameter $\tilde{C}$ defined in Eq.~\eqref{eq:tld_C}
as a function of the number of atoms $N$. The transition to the superradiant regime is shown for three
 different values of the ratio $\beta=J/\hbar\omega_0$:
   (a)~$\beta=0$,
(b)~$\beta=0.5$, and (c)~$\beta=0.9$.}
  \label{cor}
\end{figure*}

In Fig.~\ref{cor} we present the results for $N$ dependence of the
order parameter
\begin{equation}
\label{eq:tld_C} C\approx\tilde{C}=\lim_{T\rightarrow\infty}
\frac{1}{T} \int_0^Tdt \frac {\sum_j^N\sum_{i\neq
j}^N\left(\Gamma_i^3\sigma_i^+\sigma^-_j+\sigma^+_j\sigma_i^-\Gamma^3_i\right)}{\sum_{i}^N
\left(1+2\sigma_i^z\right)\Gamma_i^3}
\end{equation}
computed by solving the system~\eqref{MF_F} numerically at
different values of the interaction parameter
$\beta=J/\hbar\omega_0$. The curves show that the critical number
of atoms and the supperradiance transition are almost insensible
to the interatomic interaction. So, similar to the case of
non-interacting atoms, we have two qualitatively different regimes
of relaxation: the incoherent regime at $N<N_c$ and the regime of
superradiance at $N>N_c$.

\subsection{Collective pulse relaxation}
\label{supr-spont}

\begin{figure*}[!tbh]
\centering
 \resizebox{100mm}{!}{\includegraphics{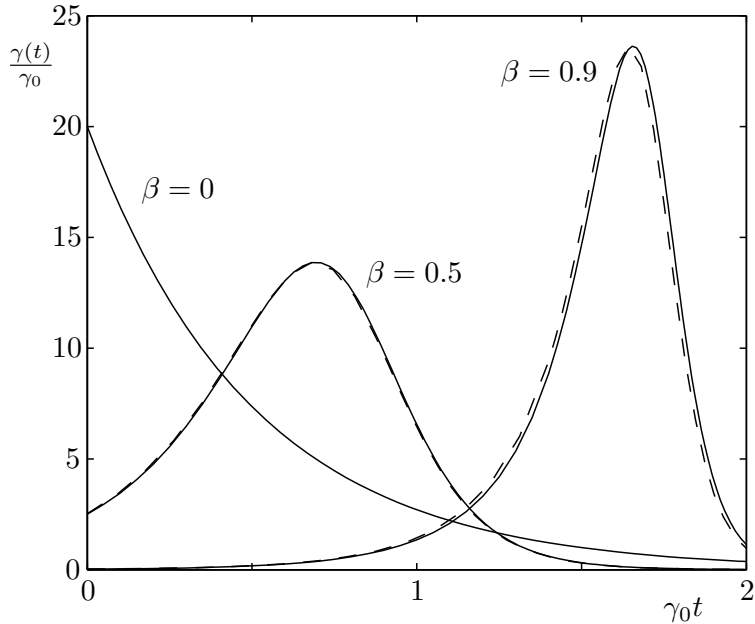}}
 \caption{Relaxation rate $\gamma(t)$ defined in Eq.~\eqref{decay-rate}
as a function of time for 10 interacting atoms
 at different values of the interaction parameter
$\beta=J/\hbar\omega_0$. Solid lines represent the numerical
results computed by solving the system~\eqref{ME1}. Dashed line is
computed from the solution of Eq.~\eqref{colrel}. }
  \label{n10}
\end{figure*}

Let us consider decay of initially inverted system of interacting
atoms at subcritical number of the atoms $N<N_c$ in more details.
Dependence of the relaxation rate on time calculated by solving
Eq.~\eqref{MF_F} with the initial conditions $\langle
S^z_n\rangle(0)=1/2$ is shown in Fig.~\ref{n10}.

Referring to Fig.~\ref{n10}, it is seen that, in the absence of
interaction ($J=0$ and $\hat\Gamma_i\equiv1$), relaxation occurs
as an incoherent spontaneous decay of excited atoms. This regime
is characterized by a monotonic exponential decrease of the
relaxation rate.

When the interaction parameter $\beta$ increases, spontaneous
processes are suppressed and the time dependence of the relaxation
rate reveals a non-monotonic behavior with a pronounced peak (see
Fig.~\ref{n10}).  This is the regime of collective pulse
relaxation induced by the direct interatomic interaction.

Collective pulse relaxation can be easily described using the mean
field approximation for decoupling of correlations. To this end we assume that, for small number
of atoms, the relaxation rate~\eqref{occup} is predominantly
determined by the first term (incoherent part) describing
incoherent mechanism of decay. So, the equation~\eqref{occup}
takes the simplified form:

\begin{equation}
\label{Sz} \frac{d}{dt}\langle
S^z_n\rangle\approx
-\gamma_0\langle(1+2S^z_n)\hat\Gamma^3_n\rangle\approx-\gamma_0\left(1+2\langle S^z_n\rangle\right)\langle\hat\Gamma_n^3\rangle
\end{equation}

We can now substitute the ansatz
\begin{equation}
  \label{eq:mean-field}
 \langle S^z_n\rangle
=\langle\tilde{S}^z\rangle+\delta\langle\tilde{S}^z_n\rangle,
\end{equation}
assuming that $\langle\tilde{S}^z\rangle\gg\delta\langle\tilde{S}^z_n\rangle$,
into Eq.~\eqref{Sz} and use the relation~\eqref{Gamma3_av} to
derive the equation governing dynamics of atomic subsystem
\begin{equation}
\label{colrel} \frac{d\langle\tilde{S}^z\rangle}{dt}
=-\gamma_0(1+2\langle\tilde{S}^z\rangle)
\langle\Gamma^3\rangle,~~~\langle\tilde{S}^z\rangle(0)=\frac{1}{2},
\end{equation}
where
\begin{equation}\label{gammamf}
\langle\Gamma^3\rangle\approx1-2\beta\left(3+\beta^2\right)\langle\tilde{S}^z\rangle+
\frac{3}{2}\beta^2\left(1+4\langle\tilde{S}^z\rangle^2\right).
\end{equation}

It can be shown that the mean field solution
$\langle\tilde{S}^z\rangle(t)$ is a step-like function of time and
is stable at
$|\langle\tilde{S}^z\rangle|\gg|\delta\langle\tilde{S}^z_n\rangle|$.
The curves calculated in the mean field approximation for the
relaxation rate
\begin{equation*}
\gamma(t)=-\sum_n\frac{d\langle
S^z_n\rangle}{dt}\approx-N\frac{d\langle\tilde{S}^z\rangle}{dt},
\end{equation*}
are shown in Fig.~\ref{n10} as dashed lines. It can be seen that
the results are in excellent agreement with the data of numerical
analysis.

It should be noted that the peak intensity is proportional to the
number of atoms $N$ which is a consequence of a short-range
character of interatomic interaction. By contrast, for the regime
of superradiance, the intensity is typically proportional to
$N^2$.

For long-range interaction, the $N$ dependence of the peak
intensity may change.  We demonstrate this effect for the
interaction of the form
$H_A=\hbar\omega_0\sum_iS^z_i-J\sum_{i,j,i\neq j}S^z_iS^z_j$.

In this case the expression~\eqref{Gamma} for the operator
$\hat\Gamma_i$ has to be replaced by the relation
$\hat\Gamma_i=1-\beta\sum_{j,j\neq i}S^z_j$ that gives an
additional $N$ dependent factor.  The modified expression for the
relaxation rate is given by
\begin{eqnarray}
\label{long-range}
\gamma(t)\approx-N\frac{d\langle\tilde{S}^z\rangle}{dt}&=&\gamma_0N(1+
2\langle\tilde{S}^z\rangle)\Bigl\{1+\frac{3}{4}(N-1)\beta^2-(N-1)[3\beta+
(N-1)\frac{\beta^3}{2}]\langle\tilde{S}^z\rangle
\nonumber\\
&+&3\beta^2(N-1)(N-2)\langle\tilde{S}^z\rangle^2-\beta^3(N-1)(N-2)(N-3)\langle\tilde{S}^z\rangle^3\Bigl\}.
\end{eqnarray}
Since the relaxation rate $\gamma(t)$ reaches the peak at
negligibly small $\langle\tilde S^z\rangle$, $\langle\tilde
S^z\rangle\approx 0$, from Eq.~\eqref{long-range} its intensity
can be estimated to be  an increasing function of $\beta$
proportional to $N^2$:
\begin{equation*}
\gamma_{\max}\approx
\gamma_0N\left[1+\frac{3}{4}(N-1)\beta^2\right]\propto N^2.
\end{equation*}

So, we have shown that the direct Ising-type interaction has a
synchronizing effect on the system behavior.  The result is that
the regime of incoherent spontaneous decay changes to the regime
of collective pulse relaxation characterized by an increase in the
radiation time.  In the subsequent section we discuss similar
effects for the regime of superradiance.

\subsection{Enhancement of superradiance}
\label{subsec:enhanc-super}

\begin{figure*}[!tbh]
\centering
 \resizebox{100mm}{!}{\includegraphics{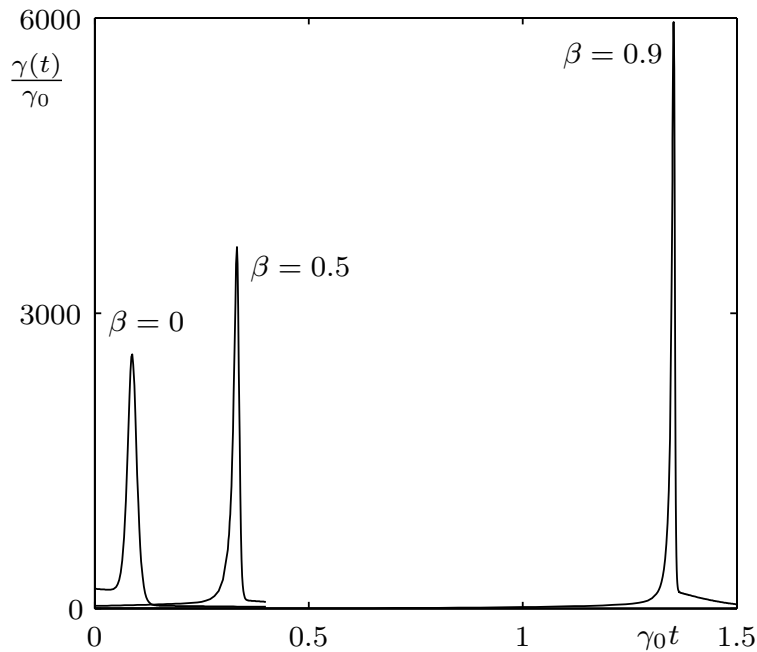}}
 \caption{Relaxation rate $\gamma(t)$ defined in
   Eq.~\eqref{decay-rate} as a function of time
for 100 interacting atoms at different values of the interaction
parameter $\beta=J/\hbar\omega_0$ ($2\gamma_0$ is the spontaneous
decay rate for an isolated atom). The curves are computed by
solving the system~\eqref{ME1} numerically. }
  \label{n100}
\end{figure*}

As it was previously discussed in
Sec.~\ref{subsec:order-parameter}, the effect of superradiance
dominates the regime of relaxation of the atomic subsystem at
supercritical values of the number of atoms, $N>N_c$.
Figure~\ref{n100} presents the time dependence of the relaxation
rate for $N=100$ at various values of the interaction parameter
$\beta$.

At $\beta=0$, the peak with the intensity proportional to $N^2$
corresponds to the superradiance effect in the system of
non-interacting atoms.  In this system interatomic correlations
are solely due to interaction between atoms and electromagnetic
field.

Synchronizing effect of the Ising-type interaction characterized
by the parameter $\beta$ manifests itself in enhancement of
superradiance and an increase in the peak intensity.

The effect of enhancement can be analyzed in the mean field
approximation by applying the procedure described in the previous
section to the coherent part (the second term) of the
relation~\eqref{occup}. So, the equation for the average of atomic
population $\langle \tilde S^z\rangle$ is now given by
\begin{equation}
\label{eq:Sz_coh} \frac{d\langle \tilde
S^z\rangle}{dt}\approx-\gamma_0N\left(\frac{3}{2}- 2{\langle \tilde
S^z\rangle^2}\right)\langle\Gamma^3\rangle,
\end{equation}
where $\langle\Gamma^3\rangle$ is given in Eq.~\eqref{gammamf} and
the identity (the conservation law for the pseudo-spin)
$\frac{1}{2}\left(S^+S^-+S^-S^+\right)+\left(S^z\right)^2=\frac{3}{4}$
is used.

The peak of the relaxation rate $\gamma(t)$ is at $\langle\tilde
S^z\rangle\approx 0$ and Eq.~\eqref{eq:Sz_coh} combined with the
relation~\eqref{gammamf} provide an estimate for the gain factor $\langle\Gamma^3\rangle\vert_{\langle \tilde
  S^z\rangle=0}\propto(1+\frac{3}{2}\beta^2)$.
As for the case of $N<N_c$, the interatomic interaction delays the
time of radiation.

For long-range interaction, following the lines of reasoning
presented in Sec.~\ref{supr-spont}, we derive dependence of the
relaxation rate on the number of atoms $N$
\begin{equation}
\gamma_{\max}\approx\gamma(t)\vert_{\langle\tilde S^z\rangle=0}=
\gamma_0N^2\frac{3}{2}\left[1+\frac{3}{4}(N-1)\beta^2\right]\propto N^3.
\end{equation}
This result differs from the $N$ dependence given at the end of
the previous section for the regime of collective pulse
relaxation.

\section{Regime of multiphoton relaxation}
\label{sec:multiphoton}

It is known that interaction between atoms at interatomic spacings
smaller than the wavelength may bring about multiphoton processes
in atomic system~\cite{Varada, Hettich}.  At weak interaction with
$J<\hbar\omega_0$, probability of multiphoton transitions is much
smaller than that of single-photon transitions.  So, the first
order of the perturbative expansion over the atom-field
interaction can be used to describe dynamical behavior of the
system.  At strong interaction with $J>\hbar\omega_0$,
single-photon processes are non-resonant and multi-photon
transitions start to play an increasingly important role. This is
the so-called regime of multiphoton relaxation.

\begin{figure*}[!tbh]
\centering
 \resizebox{100mm}{!}{\includegraphics{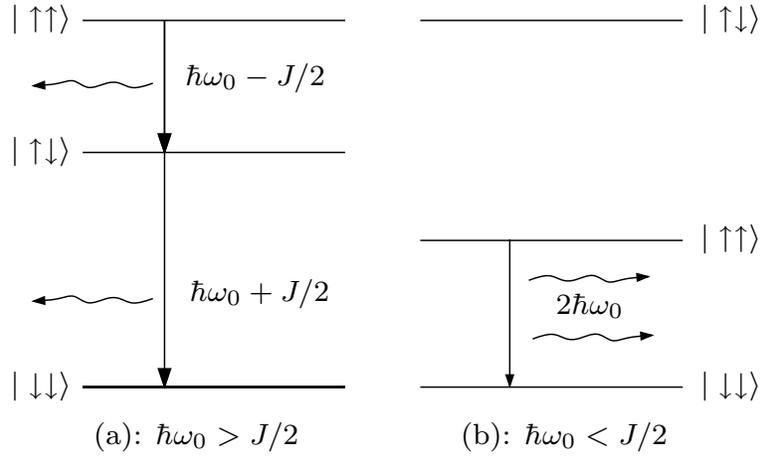}}
 \caption{Energy spectrum of two interacting atoms with
   $H_A=\hbar\omega_0\left(S^z_1+S^z_2\right)-JS^z_1S^z_2$.}
  \label{spectra2}
\end{figure*}

In order to illustrate how this regime may occur we qualitatively consider a
model system of two two-level atoms with the Hamiltonian
$H_A=\hbar\omega_0\left(S^z_1+S^z_2\right)-JS^z_1S^z_2$. Its
energy spectrum is schematically represented in
Fig.~\ref{spectra2}.

If the Hamiltonian of atom-field interaction has the form of a sum
$H_{int}=H_+e^{i\omega t}+H_-e^{-i\omega t}$, where $H_+$ ($H_-$)
describes photon emission (absorption), then the second order
perturbative expression for the probability of two-photon
transition $\vert
0\rangle\vert\uparrow\uparrow\rangle\rightarrow\vert
2\rangle\vert\downarrow\downarrow\rangle$ is given by
\begin{equation}
\label{two-phot-tr}
W^{(2)}=\frac{2\pi}{\hbar^2}\left|\frac{\langle\downarrow\downarrow|H_+
|\uparrow\downarrow\rangle\langle\uparrow\downarrow|H_+|\uparrow\uparrow\rangle}
{\hbar\omega_0(1-\beta/2)-\hbar\omega}+\frac{\langle\downarrow\downarrow|H_+
|\downarrow\uparrow\rangle\langle\downarrow\uparrow|H_+
|\uparrow\uparrow\rangle}{\hbar\omega_0(1-\beta/2)-\hbar\omega}\right|^2\delta(2\omega_0-2\omega).
\end{equation}

For strong interaction with $J>\hbar\omega_0$, the energy level of
one-particle excited states, $\vert\uparrow\downarrow\rangle$ and
$\vert\downarrow\uparrow\rangle$, is higher than the energy of the
state $\vert\uparrow\uparrow\rangle$ (see Fig.~\ref{spectra2}b),
so that one-photon relaxation of the two-particle excited state is
excluded by the energy conservation law.  As a consequence,
relaxation occurs as a two-photon process.

Probability of multiphoton transitions for $N$ excited atoms in
free space is small because its magnitude is determined by $N$-th
order term of perturbative expansion over the atom-field coupling
constant $g_{\mathbf{k}s,i}$.  So, we arrive at the conclusion
that the effect of interatomic interaction can be the formation of
long living excited states in the system.

This conclusion, however, is not strictly valid if the density of
electromagnetic modes has a singularity near the energy of atomic
transition, $\hbar\omega_0$. An example is a high quality resonant
cavity.  In this case, dynamical behavior of the atom-field system
is characterized by Rabi oscillations involving many-particle and
multiphoton states~\cite{Agarwal04}.

We demonstrate this effect for the two atomic system embedded into
a high-Q single-mode resonant cavity.  The Hamiltonian of the
model is
\begin{equation}
\label{eq:two-atoms} H=\hbar\omega_0
a^+a+\hbar\omega_0\left(S^z_1+S^z_2\right)- JS^z_1S^z_2+\hbar g
\left[a\left(S^+_1+S^+_2\right)+a^+\left(S^-_1+S^-_2\right)\right],
\end{equation}
where $a(a^+)$ is the photon annihilation (creation) operator for
the cavity mode, $g$ is the coupling constant of the interaction
between atoms and cavity mode.

The model~\eqref{eq:two-atoms} is exactly solvable and the wave
function for the system initially prepared in the state
$|\psi(0)\rangle=|\uparrow\uparrow,n\rangle$, where $n$ is the
number of photons, can be written in the explicit form:
\begin{eqnarray}
\label{psi}
|\psi(t)\rangle&=&e^{-i\omega_0(n+1)t}\Biggl\{\left[\frac{n+2}{2n+3}e^{iJ't}+g^2\frac{n+1}{D}
\left(\frac{e^{iDt}}{D-J'}+\frac{e^{-iDt}}{D+J'}\right)\right]|\uparrow\uparrow,n\rangle+
\nonumber\\
&+&\sqrt{(n+1)(n+2)}\left[\frac{g^2}{D}\left(\frac{e^{iDt}}{D-J'}+\frac{e^{-iDt}}{D+J'}\right)-
\frac{e^{iJ't}}{2n+3}\right]|\downarrow\downarrow,n+2\rangle-
\nonumber\\
&-&i\frac{g\sqrt{n+1}}{D}\sin
Dt\left(|\downarrow\uparrow,n+1\rangle+|\uparrow\downarrow,n+1\rangle\right)\Biggr\},
\end{eqnarray}
where
 \begin{equation*}
 J'=J/4\hbar,~~~D=\sqrt{(J')^2+2g^2(2n+3)}.
 \end{equation*}

 In the absence of the interatomic interaction ($J=0$), the amplitudes
 of one- and two-photon processes are of the same order of magnitude.
 But, at strong interaction with $J\gg \hbar g\sqrt{n}$, the two-particle
 two-photon amplitudes dominate and the wave function~\eqref{psi}
can be approximated as follows
\begin{eqnarray}
\label{psill} |\psi(t)\rangle&\approx&
e^{-i(\omega_0(n+1)-J'-\Delta)t}\Biggl\{\left[\frac{n+1}{2n+3}e^{i\Delta
t}+ \frac{n+2}{2n+3}e^{-i\Delta
t}\right]|\uparrow\uparrow,n\rangle+
\nonumber\\
&+&2i\frac{\sqrt{(n+1)(n+2)}}{2n+3}\sin\Delta
t|\downarrow\downarrow,n+2\rangle\Biggr\},
\end{eqnarray}
where
 \[
\Delta=\frac{g^2}{2J'}\left(2n+3\right),
 \]
and for $n\gg1$
\[
|\psi(t)\rangle\approx
e^{-i(\omega_0n-J')t}\left(\cos\Delta t|\uparrow\uparrow,n\rangle+i\sin\Delta
t|\downarrow\downarrow,n+2\rangle\right).
\]
 This means the build-up of two-atomic two-photon Rabi oscillations
 and can be used to generate a Schr\"{o}dinger cat-like entangled
 atom-field state.

Thus, interparticle interaction may give rise to the regime of
many-particle multiphoton dynamics.  When interaction is strong as
compared with the energy of atomic transition, inhibition of
one-photon processes in the system is accompanied by transition
from the regime of cooperative pulse radiation (superradiance) to
the generation of Fock state of light.  In a cavity, multiphoton
dynamical effects come into play under the condition $J\gg \hbar
g\sqrt{n}$, so that the level of intermediate energy is
essentially shifted away from one-photon resonance (see
Fig.~\ref{spectra2}).

\section{Discussion and conclusions}
\label{sec:disc-concl}

We studied effects of direct interatomic interaction on collective
processes in atom-photon dynamics using, as an example, a simple
model of two-level atoms with Ising interaction of ferromagnetic
type.  We have found that this interaction influences radiation
processes of atomic ensemble acting as an additional synchronizing
factor.

For weakly interacting atoms at $J<\hbar\omega_0$, we have shown
that interatomic interaction results in inhibition of incoherent
spontaneous decay of atoms and dynamical behavior of the system is
governed by the regime of collective pulse relaxation.  This
regime, though it bears a resemblance to superradiance, has
nothing to do with the effect of phase synchronization induced by
fluctuations of the electromagnetic field.  For a example, in
solid state structures, collective pulse relaxation is caused by
inelastic exciton-exciton scattering and is characterized by
quadratic dependence of the radiation peak on the number of
particles~\cite{Salvan,Hvam,Kondo}.

We have also found that interaction induced synchronization
enhances superradiance and can be responsible for anomalous
dependence of the radiation peak on the number of particles.  In
the presence of interparticle interaction collective pulse
radiation and enhanced superradiance are both characterized by an
increase in the delay time of emission.

At the end of Sec.~\ref{sec:weak-interaction}, we have pointed out
that when the excited state is not fully inverted its relaxation
can be determined by transitions with frequencies equal to the
atomic resonant frequency $\omega_0$.  If we consider non-excited
atoms of the initial state as defects of the atomic chain, such
relaxation scheme can be referred to as the solitonic mechanism of
relaxation. The solitonic mechanism implies that in the course of
relaxation atoms undergo transition to the ground state
successively one after another. So, such behavior can be
interpreted as a defect motion.  More generally, since the system
of equations~\eqref{MF_F} governing dynamics of atoms at
$J<\hbar\omega_0$ is similar in structure to the Volterra system
and the Toda lattice~\cite{Toda}, it might be expected that the
system posses soliton-like solutions.

\begin{figure*}[!tbh]
\centering
 \resizebox{140mm}{!}{\includegraphics{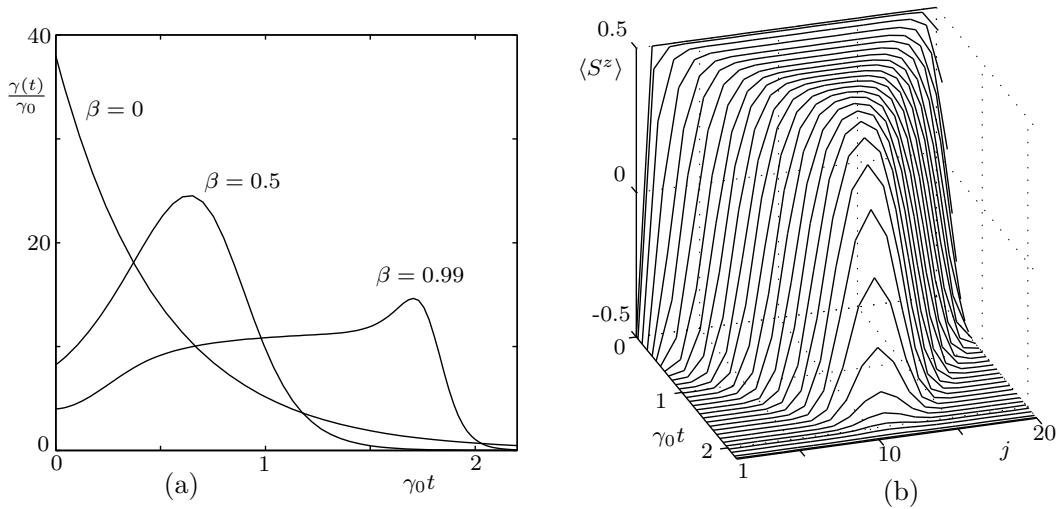}}
 \caption{(a)~Relaxation rate $\gamma(t)$ as a function of time
at different values of the interaction parameter $\beta$.
(b)~Averaged population of $i$th atom $\langle S^z_j\rangle$ as a
function of time and atomic number $i$ at $\beta=0.99$. The curves
are computed by numerically solving Eq.~\eqref{colrel} for a ring of 20
atoms. One atom is initially in the ground state representing a
point defect in the atomic ring. At sufficiently strong
interaction, relaxation of the atomic ring clearly demonstrate the
solitonic mechanism when atoms undergo transition to the ground
state consecutively one after another. }
  \label{soliton}
\end{figure*}

Figure~\ref{soliton} presents the numerical results for the excited
state with one initially non-excited atom.
The number of atoms is small, $N<N_c$, and
the relaxation rate is computed from the mean field
equation~\eqref{colrel}.

Referring to Fig.~\ref{soliton}, it is seen that relaxation of the
atomic ring can be described as motion of the defect (Bloch
wall). When the interaction parameter $\beta$ approaches the limit
of strong interaction, the relaxation rate $\gamma(t)$ assumes the
kink-like form and is determined by defect velocity.  The peak is
caused by collective pulse relaxation indicating that the
retardation time of emission is too short for the defect to travel
through the ring.

However, it should be noted that we have neglected the
dipole-dipole interaction by considering permutationally invariant
atomic configurations, the approximation that is not valid in the presence of defect in the system.  This interaction may have a destructive
effect on the solitonic mechanism of relaxation.

By contrast to the case of weakly interacting atoms where dynamics
of the atomic subsystem is governed by one-particle one-photon
transitions, at strong interatomic interaction with
$J\gg\hbar\omega_0$, these transitions appear to be excluded.  In
this case multiphoton transitions will determine relaxation of the
excited atomic subsystem.  This is what we called the regime of
multiphoton relaxation.

The regime is characterized by transition from generation of
superradiant pulse to generation of Fock quantum state of light.
We have used a simple model of two atoms in a high-Q single mode
cavity to show that such transition is accompanied by Rabi
oscillations involving many-atom multiphoton states.  In other
words, it means generating many-particle entangled atom-field
state~\cite{Pittman}.

Interestingly, transition to multiphoton dynamics is analogous to
the Mott-insulator quantum phase transition in optically trapped
atomic systems where Fock state characterizing the number of
localized atoms is formed~\cite{Greiner1,Greiner2}.  In our case,
the energy of atomic transition $\hbar\omega_0$ and the
interatomic coupling $J$ play the role of kinetic and potential
energy, respectively.  So, the transition to generation of the Fock
state of light takes place when the potential energy $J$ becomes
greater than $\hbar\omega_0$.

\begin{figure*}[!tbh]
\centering
 \resizebox{100mm}{!}{\includegraphics{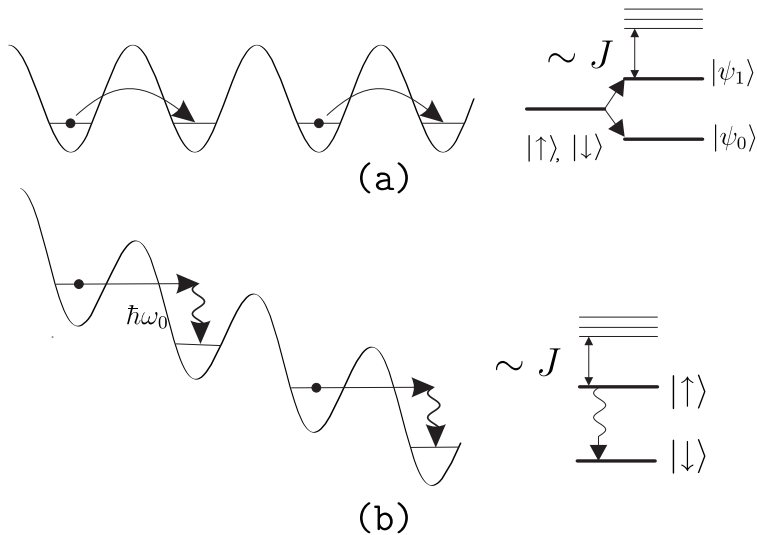}}
 \caption{
(a) Interacting particles embedded into potential lattice in the case
of half-filling. Ground state is double degenerate corresponding to
the particles localized in either odd or even wells
($\vert\uparrow\rangle=\vert\uparrow\uparrow...\uparrow\rangle$ and
$\vert\downarrow\rangle=\vert\downarrow\downarrow...\downarrow\rangle$).
Tunneling transitions lift degeneracy so that two lowest levels
correspond to the many-particles entangled states
$\vert\psi_{0,1}\rangle\approx\frac{1}{\sqrt{2}}\left(\vert\uparrow\rangle\pm\vert\downarrow\rangle\right)$.
Excited states are separated by the gap of the width $\propto J$. (b) For
energetically non-equivalent wells, tunneling transitions are
inelastic. For $J>>\hbar\omega_0$, they can be accompanied by
multiphoton processes.
}
  \label{lattice}
\end{figure*}

In lattice atomic systems with inelastic tunneling transitions between
neighboring wells, the Mott or Peierls transitions and the transition
to multiphoton relaxation can be related to each other.
We illustrate such a possibility by the simple example of an
one-dimensional periodic chain of potential wells, Fig.~\ref{lattice}.

In the half-filling
case, when there are half as many atoms as there are wells and the
energy of repulsion between neighboring particles $J$ is much larger
then the hopping energy of tunneling between neighboring wells
$\varepsilon_t$, the ground state for identical wells should be
degenerate in energy, which corresponds to the atoms localized in
either odd or even wells, see Fig.~\ref{lattice}a. Such ordering of
particles is similar to the Wigner crystal or the  charge density waves
observed in low-dimensional conductors~\cite{Gruner,Zaitsev}.
Possibility of the density waves Mott-insulator phase for atoms
embedded into an optical lattice was also discussed
in Ref.~\cite{Illuminati}.
These ordered states are separated from the excited
ones by the gap of width $\sim~J$.

Local tunneling transitions may lift this degeneracy of the ground
state.  In this case, at $J>\varepsilon_t$, the ground state can
be described, at least for a short chain, as a many-particle entangled state of the
Schr\"odinger cat type, i.e.,
$\approx\frac{1}{\sqrt{2}}\left(\vert\uparrow\uparrow...
\uparrow\rangle\pm\vert\downarrow\downarrow...\downarrow\rangle\right)$
(the so-called GHZ state~\cite{GHZ}),
associated with many-particle tunneling oscillations between odd
and even wells, (see, e.g., Ref.~\cite{Lukyanets}),
which is similar to the tunneling creep of a charge density waves~\cite{Zaitsev}.

If the wells
are energetically inequivalent, Fig.~\ref{lattice}b,
the tunneling many-particle transitions from
the excited state $\vert\uparrow\rangle=\vert\uparrow\uparrow...\uparrow\rangle$ are
inelastic and can be accompanied by multiphoton processes or
the cooperative dynamic regimes discussed above.

But our model is oversimplified and, strictly speaking,
cannot be applied to inelastic tunneling transitions
of atoms between energetically different wells.  The Hamiltonian
of atoms $H_A$ need to be modified to take into account
overlapping of the particle wave function in the neighboring
well.  This requires additional terms proportional to $S^x$.
Symmetry of interaction between particles and field $E$ also has
to be of more complicated form $H_{int}\propto
E(d_0+d_xS^x+d_zS^z)$.

Another limitation of our model is neglecting multipole
transitions that, in the case of strong interaction, may compete
with multiphoton transitions~\cite{DAN, Politi}.

Among important omissions in this paper are
spectral characteristics of the radiation. In our
case there are only two modes with the frequencies
$\hbar\omega_0\pm J$.
We also have not discussed superradiant and subradiant Dicke states.
These have the standard form because the Ising-type interparticle
interaction makes the atomic spectrum non-equidistant but does not
affect symmetry of these states.

\begin{acknowledgments}
Authors thank Alexei D. Kiselev for careful reading of the manuscript
and useful suggestions, and P. M. Tomchuk for helpful
discussions.
\end{acknowledgments}

\appendix

\section{The Born-Markov approximation}
\label{Ap-Born-Markov}

In this section we
concentrate on
the case of weakly interacting atoms with $J<\hbar\omega_0$
and describe the dissipative dynamics of the atomic
subsystem using the Born approximation.
To this end we apply to the master equation~\eqref{ME_gen_form}
the standard technique  of elimination of bosonic
variables~\cite{Carmichael}.

For simplicity, we consider the zero temperature
case and choose an initial state with no correlations between the
atomic subsystem and the vacuum field.
So,  the density operator can be taken in the factorized form~\eqref{DM_B}.

Substituting Eq.~\eqref{DM_B} into the equation~\eqref{ME_gen_form}
and taking the trace over the field variables, we obtain
\begin{eqnarray}
\label{ME_born_trace_int}
\frac{d}{dt}\tilde{\rho}(t)&=&\eta\sum_{i,j}\int_0^t
d\tau\int_0^\infty{d\omega}\omega^3
F_{ij}(\omega)\Bigl\{e^{i\omega\tau}\Biglb([R_j(t),\tilde{\rho}(t-\tau)S_i^+e^{i{\omega_0}\hat{\Gamma_i}(t-\tau)}]+
\nonumber\\
&+&
[R_j(t),\tilde{\rho}(t-\tau)S_i^-e^{-i{\omega_0}\hat{\Gamma_i}
(t-\tau)}]\Bigrb)+H.c.\Bigr\},
\end{eqnarray}
where the sum over $\mathbf{k}s$ is transformed into the
integral form~\cite{Louisell}
\begin{equation}
  \sum_{\mathbf{k}s}\longrightarrow\frac{V}{(2\pi{c})^3}\int_0^\infty{d\omega}\omega^2
\int_0^\pi{d\theta}\sin\theta\int_0^{2\pi}d\varphi\sum_s,
\end{equation}
and the functions $R_j(t)$, $F_{ij}(\omega)$, $\tilde\rho(t)$,
$\eta$ are described by Eqs.~\eqref{rho-int}-\eqref{eta}.

The next step is to employ the Markov approximation.
This approximation implies that a
reservoir relaxation time is much shorter than a time-scale of the
atomic subsystem evolution, so that
$\tilde{\rho}(t-\tau)$ can be replaced by $\tilde{\rho}(t)$
in the right hand side of Eq.~\eqref{ME_born_trace_int}
and the upper limit of the integral can be extended to infinity.

For $J<\hbar\omega_0$, the eigenvalues of the
operators $\hat\Gamma_i$ are all positive.
Using the relation
\begin{equation}
\label{int}
\lim_{t\rightarrow\infty}\int_0^td\tau{e^{\pm
ix\tau}}=\pi\delta(x)\pm i\frac{\mathcal{P}}{x},
\end{equation}
where the symbol $\mathcal{P}$ stands for
the Cauchy principal value of the integral,
we can perform the integral in Eq.~\eqref{ME_born_trace_int} and
derive Eq.~\eqref{ME_B-M_1}.
We have used the operator functions
defined on the eigenvalues of $\hat\Gamma_i$ so that if
$\hat\Gamma_i\vert\psi\rangle=\Gamma_i\vert\psi\rangle$ then, e.g.,
$\delta(\omega-\omega_0\Gamma_i)\vert\psi\rangle
=\delta(\omega-\omega_0\hat\Gamma_i)\vert\psi\rangle$.

This result can also be obtained directly by representing
the terms $\exp(\pm i\omega_0\hat\Gamma_i\tau)$
that enter Eq.~\eqref{ME_born_trace_int} as follows
\begin{equation}
\label{exp}
e^{\pm{i\omega_0\hat{\Gamma}_i}\tau}=\left(1-\hat{U}_i\right)e^{\pm{i\omega_0}\tau}+\frac{1}{2}\left(\hat{U}_i+
\hat{\Omega}_i\right)e^{\pm{i\omega_0(1-\beta)}\tau}+\frac{1}{2}\left(\hat{U}_i-
\hat{\Omega}_i\right)e^{\pm{i\omega_0(1+\beta)}\tau},
\end{equation}
where
\begin{equation}
\hat{U}_i=\frac{1}{2}\left(1+4S_{i+1}^zS_{i-1}^z\right),~~~\hat{\Omega}_i=S_{i+1}^z+S_{i-1}^z.
\end{equation}

The principal values of the integrals in~\eqref{ME_B-M_1} with $i\neq{j}$ can be estimated as:
\begin{subequations}
\label{d-d-accur}
\begin{eqnarray}
&&\mathcal{P}\int_0^\infty{d\omega}\omega^3\frac{F_{ij}(\omega)}{\omega+
\omega_0\hat{\Gamma}_i}=\frac{3}{2}\left(\omega_0\hat\Gamma_i\right)^3
\Biggl\{\left[\frac{1-(\bar{\mathbf{d}}\cdot\bar{\mathbf{r}}_{ij})^2}{k_0\hat\Gamma_ir_{ij}}-
\frac{1-3(\bar{\mathbf{d}}\cdot\bar{\mathbf{r}}_{ij})^2}{(k_0\hat\Gamma_ir_{ij})^3}\right]A(k_0\hat\Gamma_ir_{ij})+
\nonumber
\\
&&+\frac{[1-3(\bar{\mathbf{d}}\cdot\bar{\mathbf{r}}_{ij})^2]B(k_0\hat\Gamma_ir_{ij})-
[1-(\bar{\mathbf{d}}\cdot\bar{\mathbf{r}}_{ij})^2]}{(k_0\hat\Gamma_ir_{ij})^2}\Biggr\},
\\
\nonumber\\
&&\mathcal{P}\int_0^\infty{d\omega}\omega^3\frac{F_{ij}(\omega)}{\omega-\omega_0\hat{\Gamma}_i}
=\frac{3\pi}{2}\left(\omega_0\hat\Gamma_i\right)^3\Biggl\{[1-(\bar{\mathbf{d}}\cdot\bar{\mathbf{r}}_{ij})^2]
\frac{\cos(k_0\hat\Gamma_ir_{ij})}{k_0\hat\Gamma_ir_{ij}}-
\nonumber\\
&&-[1-3(\bar{\mathbf{d}}\cdot\bar{\mathbf{r}}_{ij})^2]\left[\frac{\sin(k_0\hat\Gamma_ir_{ij})}{(k_0\hat\Gamma_ir_{ij})^2}
+\frac{\cos(k_0\hat\Gamma_ir_{ij})}{(k_0\hat\Gamma_ir_{ij})^3}\right]\Biggr\}
-\mathcal{P}\int_0^\infty{d\omega}\omega^3\frac{F_{ij}(\omega)}{\omega+\omega_0\hat{\Gamma}_i},
\end{eqnarray}
\end{subequations}
where $k_0=\omega_0/c$,
\begin{eqnarray*}
&&A(x)=\sin(x)\ci(x)-\cos(x)\si(x),\\
&&B(x)=\sin(x)\si(x)+\cos(x)\ci(x),~~~x>0,
\end{eqnarray*}
and $\si(x)=\int_\infty^x\frac{\sin t}{t}dt$ and
$\ci(x)=\int_\infty^x\frac{\cos t}{t}dt$~\cite{Bateman}.

In the limit  $r_{ij}\omega/c\rightarrow 1$,
Eq.~\eqref{d-d-accur} assumes the asymptotical
form~\eqref{d-d-as} coincident
with the standard expression for the
dipole-dipole interaction that does not depend on
the coupling constant of Ising interaction $J$.

\section{Two interacting atoms: exact solution of the master equation}
\label{Ap-2atBM}

For the case of two non-interacting atoms with $J=0$,
exact solution of the master equation~\eqref{ME1}
was previously obtained in Ref.~\cite{Coffey}.
It was shown that the decay of initially exited atoms demonstrates
the superradiant regime  and the dipole-dipole interaction does
not influence the cooperative behavior of atoms.

In this section we show
that  the master equation~\eqref{ME1}
for two Ising-like interacting atoms, $J\neq 0$,
can be solved along similar lines.

The Hamiltonian of two interacting atoms is given by
\begin{equation*}
H_A= \hbar\omega_0\left(S^z_1+S^z_2\right)-JS^z_1S^z_2.
\end{equation*}

In the  basis of atomic states $\vert\uparrow\uparrow\rangle$,
$\vert\uparrow\downarrow\rangle$,
$\vert\downarrow\uparrow\rangle$,
$\vert\downarrow\downarrow\rangle$ the density operator $\rho$
can be written the matrix form
\begin{equation}
\rho=\begin{pmatrix}
\rho_{11} & 0 & 0 & 0\\
0 & \rho_{22} & \rho_{23} & 0 \\
0 & \rho_{32} & \rho_{33} & 0 \\
0 & 0 & 0 & \rho_{44}
\end{pmatrix}
\end{equation}
and Eq.~\eqref{ME1} reduces to the system
\begin{eqnarray}
&&\dot\rho_{11}=-4m\rho_{11}, \label{2at-syst} \nonumber\\
&&\dot{x}_0=4m\rho_{11}-2n(x_0+x_1), \nonumber\\
&&\dot{x}_1=4m\rho_{11}-2n(x_1+x_0), \\
&&\dot{x}_2=-2nx_2+i2\Omega x_3, \nonumber\\
&&\dot{x}_3=i2\Omega x_2-2nx_3, \nonumber\\
&&\dot{\rho}_{44}=2n(x_0+x_1),\nonumber
\end{eqnarray}
where $x_0=\rho_{22}+\rho_{33}$, $x_1=\rho_{23}+\rho_{32}$,
$x_2=\rho_{23}-\rho_{32}$, $x_3=\rho_{22}-\rho_{33}$,
$n=\gamma_0(1+\beta/2)^3$, $m=\gamma_0(1-\beta/2)^3$ and
$\Omega_{12}=\Omega_{21}=\Omega$ is the constant of the
dipole-dipole interaction.

For the initial conditions $\rho_{11}(0)=1$,
$\rho_{22}(0)=\rho_{23}(0)=\rho_{32}(0)=\rho_{33}(0)=\rho_{44}(0)=0$
describing two initially excited atoms,
it is easy to write down the solution of Eq.~\eqref{2at-syst}
\begin{eqnarray}
\label{solution}
&&\rho_{11}=e^{-4mt},\nonumber\\
&&x_0=x_1=\frac{m}{n-m}\left(e^{-4mt}-e^{-4nt}\right),\\
&&\rho_{44}=1+\frac{1}{m-n}\left(ne^{-4mt}-me^{-4nt}\right)\nonumber,
\end{eqnarray}
and also $x_2=x_3=0$.

It is seen that similar to the case of non-interacting atoms
the solution~\eqref{solution} and the decay rate of
the atomic subsystem
\begin{equation}
\gamma(t)\equiv-\frac{d}{dt}{Tr}\lbrace\rho(S_1^z+S_2^z)\rbrace=\dot\rho_{44}-\dot\rho_{11}
\end{equation}
are both independent of the constant of the dipole-dipole interaction $\Omega$.

\section{Elimination of rapidly oscillating variables}
\label{Ap-fast-var}

In order to eliminate from~\eqref{Bloch2} the rapidly oscillating terms
$i\alpha\langle\hat\Gamma_n\rangle\langle S^\pm_n\rangle$
we apply the
method of multitime scales~\cite{Pobedrya,Nayfeh} representing the
atomic variables $\langle \vec{S}_n\rangle$ as functions of two time
scales $\langle\vec{S}_n\rangle(\tau,\tau')$, where
$\tau'=\alpha\tau~(\alpha\gg1)$ is a characteristic time of fast
motion.
We use the following power series expansion of $\langle S^\pm_n\rangle$ and
$\langle S^z_n\rangle$ over the small parameter $\alpha^{-1}$
\begin{subequations}
\label{razl}
\begin{eqnarray}
\label{razla}
\langle
S^\pm_n\rangle(\tau,\tau')&=&\delta^\pm_n(\tau')\tilde\sigma^\pm_n(\tau,\tau')
=\delta^\pm_n(\tau')\left({\sigma^\pm_n(\tau)}+
\frac{1}{\alpha}\sigma^\pm_{n,1}(\tau)\gamma_n^\pm(\tau')+...\right),
\\
\label{razlb}
\langle S^z_n\rangle(\tau,\tau')&=&\tilde\sigma^z_n(\tau,\tau')=
\sigma^z_n(\tau)+\frac{1}{\alpha}\sigma^z_{n,1}(\tau)\gamma_n^z(\tau')+\ldots,
\end{eqnarray}
\end{subequations}
where $\gamma^\pm_n(\tau')$, $\gamma^z_n(\tau')$ and
$\delta^\pm_n(\tau')$ are rapidly oscillating periodic functions, so
that
\begin{equation*}
\overline{(\gamma_n^\pm)}_{\tau'}\equiv\frac{1}{2\pi}\int_0^{2\pi}\gamma_n^\pm(\tau')d\tau'
=0.
\end{equation*}

Taking into account that
\begin{equation}
\label{diff}
\frac{d}{d\tau}=\frac{\partial}{\partial\tau}+\alpha\frac{\partial}{\partial
\tau'}
\end{equation}
and substituting the expansion~\eqref{razla} into Eq.~\eqref{Bloch2},
 we deduce the following relation
for $\tilde\sigma^+_n$
\begin{eqnarray}
\label{eq1}
&&\left(\frac{\partial\tilde\sigma_n^+}{\partial\tau}+\alpha\frac{\partial\tilde\sigma_n^+}
{\partial\tau'}\right)\delta_n^++\alpha\tilde\sigma_n^+\frac{\partial\delta_n^+}{\partial\tau'}
=\left(i\alpha\langle\hat\Gamma_n\rangle-
\langle\hat\Gamma_n^3\rangle\right)\tilde\sigma_n^+\delta_n^++
\nonumber\\
&&+2\left(\langle\hat
E_{n+1}\rangle\tilde\sigma_{n+1}^+\delta_{n+1}^++\langle\hat
E_{n-1}\rangle\tilde\sigma_{n-1}^+\delta_{n-1}^+\right)+2\sum_{i\neq
n,n\pm1}\langle\hat
S^z_n\rangle\langle\hat\Gamma_i^3\rangle\tilde\sigma_{i}^+\delta_{i}^+.
\end{eqnarray}

In order to eliminate the imaginary term in the right hand side of
Eq.~\eqref{eq1} we choose $\delta_n^+(\tau')$ in the form
\begin{equation*}
\delta_n^+(\tau')=\exp\left(i\int_0^{\tau'}\langle\hat\Gamma_n\rangle(\tau,\xi)
d\xi\right)\approx e^{i\Gamma_n\tau'},
\end{equation*}
where $\Gamma_n$ is the zero order term of the expansion for
$\langle\hat\Gamma_n\rangle(\tau,\tau')$ given by~\eqref{Gamma_av},
to yield the equation
\begin{eqnarray}
\label{eq2}
&&\frac{\partial\tilde\sigma_n^+}{\partial\tau}+\alpha\frac{\partial\tilde\sigma_n^+}{\partial\tau'}
=-\langle\hat\Gamma_n^3\rangle\tilde\sigma_n^++
2\biglb(\langle\hat E_{n+1}\rangle\tilde\sigma_{n+1}^+
e^{i\left(\Gamma_{n+1}-\Gamma_{n}\right)\tau'}+
\nonumber
\\
&&+\langle\hat
E_{n-1}\rangle\tilde\sigma_{n-1}^+e^{i\left(\Gamma_{n-1}-\Gamma_{n}\right)\tau'}\bigrb)+2\sum_{i\neq
n,n\pm1}\langle\hat
S^z_n\rangle\langle\hat\Gamma_i^3\rangle\tilde\sigma_{i}^+e^{i\left(\Gamma_{i}-\Gamma_{n}\right)\tau'}.
\end{eqnarray}

Averaging Eq.~\eqref{eq2} over $\tau'$ and retaining only the lowest
order of the correlations,  we have
\begin{eqnarray}
\label{eq3}
\frac{d\sigma_n^+}{d\tau}=
-\Gamma_n^3\sigma_n^++2\left(E_{n+1}\sigma_{n+1}^+w_{n+1,n}+E_{n-1}\sigma_{n-1}^+w_{n-1,n}\right)+
2\sum_{i\neq n,n\pm1}\Gamma_i^3\sigma_{i}^+\sigma^z_nw_{i,n},
\end{eqnarray}
where $w_{i,j}$ is given by
\[w_{i,j}=\frac{e^{i2\pi(\Gamma_i-\Gamma_j)}-1}{i2\pi(\Gamma_i-\Gamma_j)}.\]

Equation~\eqref{eq3} is the second equation of the system~\eqref{MF_F}.
The complex conjugate of Eq.~\eqref{eq3}
gives the equation for $\sigma^-_n$.
Along the same line
averaging the equation for $\langle S^z_n\rangle$ from Eq.~\eqref{Bloch2}
provides the last equation of Eq.~\eqref{MF_F}.

\end{document}